\documentclass[lettersize,journal]{IEEEtran}
\usepackage{amsmath,amsfonts,amssymb}
\usepackage{algorithmic}
\usepackage{array}
\usepackage[caption=false,font=normalsize,labelfont=sf,textfont=sf]{subfig}
\usepackage{textcomp}
\usepackage{stfloats}
\usepackage{url}
\usepackage{verbatim}
\usepackage{graphicx}
\usepackage{booktabs}
\usepackage{amsmath}
\usepackage{bm}

\def\BibTeX{{\rm B\kern-.05em{\sc i\kern-.025em b}\kern-.08em
    T\kern-.1667em\lower.7ex\hbox{E}\kern-.125emX}}
\usepackage{balance}

\begin{document}
\title{A Synchronized Multi-IMU Wearable System for Tracking of Joint-Angles in Sports Motion Analysis With Reference-Based Validation and Dynamic Task Characterization}

\author{Thevindu~Samarasekera,
        Praveen~Rathnayaka,
        Sachintha~Adhikari,
        Tharinda~Navarathne,
        Mahela~Pandukabhaya,~\IEEEmembership{Graduate Student Member,~IEEE},
        Roshan~Godaliyadda,~\IEEEmembership{Senior Member,~IEEE},
        Parakrama~Ekanayake,~\IEEEmembership{Senior Member,~IEEE},
        Chanaka~Senanayake,
        Vijitha~Herath,~\IEEEmembership{Senior Member,~IEEE},
        and~Asela~Ratnayake.%
\thanks{Manuscript submitted June 09, 2026.}
\thanks{Thevindu Samarasekera, Praveen Rathnayaka, Sachintha Adhikari, Tharinda Navarathne, Mahela Pandukabhaya, Roshan Godaliyadda, Parakrama Ekanayake, and Vijitha Herath is with the Department of Electrical and Electronic Engineering, 
University of Peradeniya, Peradeniya, Sri Lanka. (e-mail: e19344@eng.pdn.ac.lk).}
\thanks{Chanaka Senanayake is with the Department of Manufacturing and Industrial Engineering, of Peradeniya, Peradeniya, Sri Lanka.}
\thanks{Asela Ratnayake is with the National Hospital of Kandy, Kandy, Sri Lanka.}
\thanks{The first three authors contributed equally to this work.}
}

% \author{IEEE Publication Technology Department
% \thanks{Manuscript created June, 2026; This work was developed by University of Peradeniya. This work is distributed under the \LaTeX \ Project Public License (LPPL) ( http://www.latex-project.org/ ) version 1.3. A copy of the LPPL, version 1.3, is included in the base \LaTeX \ documentation of all distributions of \LaTeX \ released 2003/12/01 or later. The opinions expressed here are entirely that of the author. No warranty is expressed or implied. User assumes all risk.}}

\markboth{IEEE Transactions on Instrumentation and Measurement}%
{IEEE Transactions on Instrumentation and Measurement}

\maketitle
\begin{abstract}
Reliable joint-angle measurement outside laboratory environments is important for posture assessment and technique analysis in sports and rehabilitation. Yet, wearable IMU systems being the most viable option in such applications remain constrained by drift, multi-sensor timing consistency, practical validation, and cost-efficiency. This paper presents a synchronized, cost-efficient, modular IMU wearable platform and an end-to-end pipeline for joint-angle estimation tailored to high-dynamic sports motion analysis. The system combines high-rate sensing with robust local logging, a Real Time Clock (RTC) - disciplined microsecond timestamping scheme for inter-node timing consistency, an indirect Kalman filter based (IKF) orientation estimator followed by relative-rotation joint-angle extraction, high-pass drift mitigation, and range normalization. Reference-based validation was performed using a standardized seated knee flexion-extension protocol, comparing the IMU-derived knee trajectory against a markerless vision reference computed from YOLOv11. The proposed method reproduced the expected 14-cycle structure and achieved low normalized error. Instrumentation performance was further assessed using a long-duration rigid-body elbow hold (2~h 12~min), yielding near-zero drift rate ($r_{\mathrm{drift}}=1.5\times10^{-6}$ deg/min) and a practical noise-limited resolution of $0.6442^\circ$. Finally, the utility of the system was demonstrated on a high-dynamic clean \& jerk task with two participants (professional national-level and amateur) performing five consecutive lifts per trial. The measured elbow and knee trajectories preserved stage-dependent signatures and fast transients for technique interpretation and enabled expertise-level discrimination via differences in transient timing and trajectory regularity. Overall, the results support the proposed system as a practical, temporally consistent wearable instrumentation approach for multi-joint technique analysis in high-dynamic settings.

\end{abstract}


% \IEEEPARstart{W}{elcome} to the updated and simplified documentation to using the IEEEtran \LaTeX \ class file. The IEEE has examined hundreds of author submissions using this package to help formulate this easy to follow guide. We will cover the most commonly used elements of a journal article. For less common elements we will refer back to the ``IEEEtran\_HOWTO.pdf''.

% This document applies to version 1.8b of IEEEtran. 

% The IEEEtran template package contains the following example files: 
% \begin{list}{}{}
% \item{bare\_jrnl.tex}
% \item{bare\_conf.tex}
% \item{bare\_jrnl\_compsoc.tex}
% \item{bare\_conf\_compsoc.tex}
% \item{bare\_jrnl\_comsoc.tex}
% \end{list}
% These are ``bare bones" templates to quickly understand the document structure.  

% It is assumed that the reader has a basic working knowledge of \LaTeX. Those who are new to \LaTeX \ are encouraged to read Tobias Oetiker's ``The Not So Short Introduction to \LaTeX '', available at: \url{http://tug.ctan.org/info/lshort/english/lshort.pdf} which provides an overview of working with \LaTeX.   

% \section{Methodology}
% \input{Methodology}

\section{Introduction}
\label{sec:intro}

Reliable joint-angle measurement outside laboratory environments is essential for posture assessment, technique analysis, and injury-risk related monitoring in sports and rehabilitation \cite{[1]Wei2024-nc}.
While optical motion-capture systems provide high accuracy, they require controlled environments, specialized infrastructure, and substantial setup effort; while privacy issues limit their scalability in real-world settings \cite{[2]s24092947}.
Wearable inertial sensors offer a low-cost and portable alternative, but practical deployment remains challenging due to long-term drift, inter-sensor synchronization requirements, and sensitivity to sensor placement. \cite{[3]s22093225}.

This work addresses these challenges by presenting a complete wearable instrumentation and processing framework for multi-joint posture estimation in dynamic tasks. 
Our contributions are:
\begin{itemize}
    \item A multi-node wearable IMU platform with robust acquisition and logging for multi-joint motion tracking.
    \item An RTC-disciplined, microsecond-resolution timestamping and synchronization strategy designed for sustained inter-node timing consistency.
    \item A signal-processing pipeline that estimates segment orientations using an indirect Kalman filter, computes joint angles from relative segment rotation, and mitigates drift using high-pass filtering with range normalization.
    \item Experimental study that includes reference-based validation on a standardized knee flexion--extension protocol. Next, a long duration rigid body test is explained for resolution analysis, and drift mitigation verification. Finally, a performance characterization on a dynamic clean \& jerk weightlifting task with professional vs amateur execution.
\end{itemize}

This article is an extended version of our previously published proceedings paper \cite{11499905}, and presents technical extensions through new measurement definitions, expanded algorithmic processing, reference-based validation experiments, long-duration instrumentation performance characterization, and dynamic-task analysis for sports motion measurement.

The remainder of this paper is organized as follows: Section~\ref{sec:related} reviews related work; Section~\ref{sec:objective_metrics} talks about the measurand and performance metrics; Section~\ref{sec:System Overview and Design Rationale} describes the system design; Section~\ref{sec:pipeline} details the measurement model and processing pipeline; Section~\ref{sec:validation_standardized} presents reference-based validation; Section~\ref{sec: sensitivity, resolution, drift} explains the resolution and drift attenuation performance; and Sections~\ref{sec:protocol_cj}--\ref{sec:perf_cj} report dynamic-task protocol and performance characterization.

\section{Related Work}
\label{sec:related}

This section summarizes the most relevant prior work and existing approaches related to our study. The literature is grouped as \textit{Reference systems for joint-angle measurement}, \textit{Wearable IMU-based joint-angle estimation}, \textit{Fusion and drift mitigation}, \textit{Multi-sensor synchronization and data integrity}, and \textit{wearable motion capture devices for sports}.

\subsection{Reference Systems for Joint-angle Measurement}

Optical marker-based motion capture remains a common laboratory reference for joint kinematics, but its accuracy is limited by soft-tissue artifact and marker placement variability, which can introduce task dependent joint-angle errors even under controlled conditions \cite{[R1.1]buchman2022can}. Recent work has also formalized marker-position uncertainty propagation into joint-angle uncertainty, helping quantify how “reference” errors depend on marker configuration and affect the subsequent joint-angle calculation \cite{[R2]scheiterer2024marker}. In parallel, validated markerless video pipelines have matured into credible kinematic reference options for certain tasks, enabling lower-cost joint-angle estimation from smartphones or small camera setups with performance characterized against laboratory systems \cite{[R3]uhlrich2023opencap}\cite{[R3.1]falisse2025marker}. 
However, these systems typically require dedicated infrastructure, careful calibration, and constrained environments, limiting their use for routine field deployment \cite{[2]s24092947}.

\subsection{Wearable IMU-based Joint-angle Estimation}

In recent years, inertial measurement unit (IMU) based wearable systems have emerged as a transformative alternative to traditional optical motion capture technologies. A growing body of high-quality research demonstrates that IMU-based systems have evolved from experimental prototypes to reliable motion capture solutions, particularly in applications such as rehabilitation monitoring, sports biomechanics, and daily activity analysis \cite{[R5]slade2021open}\cite{[R7]al2022opensense}. There are a number of recent studies that support the claim that IMU-based wearable systems can produce joint-angle results that are comparable to optical motion capture \cite{9406045}\cite{[R8]robert2020inertial}\cite{[R9]liu2019wearable}\cite{Nijmeijer2023Xsens}. However, as noted in recent work, the lack of standardization across IMU-based wearable systems continues to be a key limitation \cite{[R12]cereatti2024isb}. Another practical limitation is that IMUs may shift relative to the body segment during movement, and recent work has explicitly investigated detection and correction of such IMU movements during joint-angle estimation \cite{9758716}. Therefore, there is still a need for standardized and generalizable IMU-based joint-angle estimation methods that work reliably across devices, movements, sensor placements, and populations. Moreover, sensor disposition and redundancy are also important design considerations in wearable sensing, with recent work on EMG-IMU multimodal systems showing how sensor configurations can be optimized to reduce redundancy while preserving task-relevant information \cite{9999246}.

\subsection{Fusion and Drift Mitigation}

Sensor fusion in wearable motion capture typically combines gyroscope integration with accelerometer based inclination correction and, when available, magnetometer-based heading stabilization \cite{[R13]laidig2023vqf}\cite{11456618}. 
For IMU-only configurations, robust Kalman-filter formulations and adaptive covariance tuning have been investigated to improve roll and pitch estimation without relying on magnetometer measurements \cite{9511443} \cite{5462839}. Adaptive attitude-estimation approaches for low-cost MEMS IMUs have also been proposed to reduce attitude errors caused by dynamic acceleration and magnetic disturbances, which is particularly relevant for wearable systems operating outside controlled laboratory environments \cite{9003495}.
For magnetometer free cases, kinematic constraints (e.g., approximate 2-DoF joint models and range-of-motion constraints) can reduce long-term heading drift by exploiting anatomical coupling between adjacent segments \cite{[R14.1]olsson2020robust}. Drift mitigation can also be addressed in downstream kinematics via constrained inverse-kinematics pipelines that can detect physiologically implausible orientations and reduce long-duration drift in joint trajectories \cite{[R7]al2022opensense}. Earlier work on upper-limb inertial measurement also demonstrated that drift in wrist and elbow motion estimates can be substantially reduced using Kalman-filter-based fusion of accelerometer and gyroscope measurements \cite{5247091}. Despite some of these advancements there is still no standardized sensor fusion methods that can be used for magnetometer free systems or environments with high magnetic field interference.

\subsection{Multi-sensor Synchronization and Data Integrity} 

Accurate multi-joint kinematics and coordination analysis require consistent timing across sensor nodes, as even small inter-node timing offsets can distort event timing, phase relationships, and derived asymmetry measures, particularly in fast, impact-rich sports tasks \cite{[R17]ohara2024microsync}\cite{[R17.1]asgarian2021bluesync}. Recent wearable research has demonstrated that high-accuracy synchronization can be achieved at the application layer over Bluetooth Low Energy, with reported time-alignment errors ranging from sub-millisecond to tens/hundreds of microseconds depending on platform and method design \cite{[R18]wang2023comparison}\cite{[R18.1]verdel2023time}. More broadly, high-precision synchronization remains an active instrumentation problem in distributed measurement systems, with recent work demonstrating progressive synchronization strategies capable of very high timing accuracy \cite{10926521}. Beyond purely algorithmic timestamp exchange, learning-based Bluetooth synchronization has been shown to correct time shifts across multiple wearable nodes and support synchronized high-rate motion capture \cite{[R19]balasubramanian2023neural}\cite{[R19.1]landra2025sharktooth}. These findings highlight that the advancement of inter-device synchronization techniques is critical to unlocking the full potential of wearable IMU systems for accurate, high-fidelity multi-joint motion analysis.

\subsection{Wearable Motion Capture Devices for Sports}

Sport-specific wearable motion capture must operate under higher dynamics than many clinical gait scenarios, with rapid transitions, impacts, and large ROM that magnify sensitivity to sensor placement, synchronization errors, and drift \cite{[2]s24092947}. Recent concurrent validation studies of commercial inertial systems in sport-relevant tasks (e.g., jump-landing and change-of-direction) show strong agreement for sagittal-plane waveforms but more variable agreement/error in frontal and transverse planes, reinforcing the importance of careful calibration and interpretation of non-sagittal DOFs \cite{Nijmeijer2023Xsens}\cite{[R20.2]lloyd2024future}. Similarly, sport-specific upper-limb validations (e.g., tennis strokes) provide evidence on where wearable inertial kinematics is trustworthy and where movement-specific dynamics can degrade accuracy \cite{[R21]pedro2021concurrent}. Systematic reviews in elite sports biomechanics summarize common methodological pitfalls (reference choice, calibration, sensor placement and attachment, drift handling) and provide a framework for reporting validity/reliability in ecological running assessments \cite{[R22]mason2023wearables}. Collectively, these findings highlight that, despite inherent challenges, wearable IMU systems represent a powerful and practical tool for sport-specific motion analysis, enabling detailed kinematic assessment in dynamic, real-world environments beyond the constraints of laboratory-based methods.

\section{Measurement Objective and Performance Metrics}
\label{sec:objective_metrics}

This section defines the measurand produced by the proposed wearable instrumentation, the associated coordinate/sign conventions, and the performance metrics reported in the remainder of the manuscript.

\subsection{Measurand Definition and Derived Posture Variables}
\label{sec:measurand}

The primary measurand is the joint flexion/extension angle about the dominant hinge axis for sagittal-plane motion, estimated from the relative orientation of proximal and distal IMU-instrumented segments.
For a given joint, the proximal and distal segment orientations are denoted by $\mathbf{C}_{p}(t)$ and $\mathbf{C}_{d}(t)$, respectively.
The relative rotation is
\begin{equation}
\mathbf{C}_{\mathrm{rel}}(t) = \mathbf{C}_{p}(t)^{\top}\mathbf{C}_{d}(t),
\end{equation}
from which a scalar hinge-angle component is extracted (Section~\ref{sec:joint_angle_est}).
We report the joint-angle trajectory as $\theta(t)$, with the final drift-mitigated, range-normalized output denoted $\theta_{\mathrm{final}}(t)$.

These joint-angle time series support downstream posture and technique variables, including ROM, stage-dependent extrema/timing features, and steadiness metrics during quasi-static holds.

% \subsection{Coordinate Frames and Sign Conventions}
% \label{sec:frames_sign}

% Each IMU provides measurements in its local sensor frame $\{\mathcal{B}\}$.
% Segment orientations are estimated relative to an earth frame $\{\mathcal{E}\}$, which is gravity-referenced through accelerometer-based correction in the inertial filter.
% For a joint with proximal segment $p$ and distal segment $d$, the joint motion is represented by the relative rotation $\mathbf{C}_{\mathrm{rel}}(t)$.
% A consistent flexion direction is enforced through a joint-dependent sign convention, implemented as a scalar factor applied to the extracted hinge-angle component.

% All reported elbow and knee joint angles follow a convention in which the mapped angle increases/decreases consistently with flexion/extension across trials, and the final output is range-normalized to a physiologically meaningful interval (Section~\ref{sec:joint_angle_est}).

\subsection{Reported Performance Metrics}
\label{sec:metrics}

Performance is characterized at both the reference-validation level (standardized motion), long duration test on a rigid body, and the dynamic multi-joint task level (clean \& jerk), using the following metrics.

\paragraph*{Reference-based accuracy (standardized motion)}
When a vision-derived reference is available (Section~\ref{sec:validation_standardized}), we report:
\begin{itemize}
    \item \textbf{RMSE:} $\sqrt{\frac{1}{K}\sum_{k=1}^{K}\big(\theta_{\mathrm{IMU}}(k)-\theta_{\mathrm{ref}}(k)\big)^2}$,
    \item \textbf{MAE:} $\frac{1}{K}\sum_{k=1}^{K}\big|\theta_{\mathrm{IMU}}(k)-\theta_{\mathrm{ref}}(k)\big|$,
    \item \textbf{Bias:} $\frac{1}{K}\sum_{k=1}^{K}\big(\theta_{\mathrm{IMU}}(k)-\theta_{\mathrm{ref}}(k)\big)$,
    % \item \textbf{Cycle timing error:} peak-to-peak timing residuals across repeated cycles.
\end{itemize}

% \paragraph{Repeatability (inter-trial variability).}
% For participants with consistent execution across sessions, repeatability is summarized using trial-level feature stability across two trials (Section~\ref{sec:cj_repeatability}), including:
% (i) absolute inter-trial differences $|f_1-f_2|$ and (ii) coefficient of variation
% $\mathrm{CV}(f)=\frac{\sigma_f}{\mu_f}\times 100\%$ for selected measurement-derived features (e.g., elbow ROM and mean angle).

% \paragraph{Drift and long-duration stability.}
% Drift is characterized using baseline windows before and after a multi-lift trial (Section~\ref{sec:cj_long_drift}), reporting:
% (i) baseline mean shift $\Delta\theta_{\mathrm{drift}}$ and (ii) drift rate $r_{\mathrm{drift}}$ in deg/min.

% \paragraph{Inter-sensor synchronization consistency.}
% Multi-sensor timing consistency is assessed during dynamic motion using event-timing residual dispersion across repeated lifts (Section~\ref{sec:cj_sync}), reporting statistics of
% $\Delta t_i = t^{(a)}_i - t^{(b)}_i$ and relating results to the 100~Hz sampling period (10~ms per sample).

% \paragraph{Missing-data resilience and data integrity.}
% We report the presence/absence of data loss and its impact on downstream analysis via:
% (i) trial inclusion criteria (minimal data loss; complete set of lifts) and
% (ii) interpolation/gating steps where applicable (e.g., reference extraction).

\section{System Overview and Design Rationale}
\label{sec:System Overview and Design Rationale}
The system supports untethered, high-rate motion capture with reliable storage and tight multi-node timing. Wi-Fi enables free movement and streaming, while local logging protects against wireless dropouts. A shared time base enables accurate cross-sensor alignment for coordinated analysis.

\subsection{Implementation of Hardware System}

\subsubsection{The Sensing and Central Processing Modules}

The wearable is built around the Seeed Studio XIAO ESP32-S3 microcontroller (21 mm x 17.5 mm) with integrated Wi-Fi 802.11 b/g/n. Motion sensing uses the Bosch Sensortec BMI120 6-axis IMU, whose selectable accelerometer (±2/4/8/16 g) and gyroscope (±125 to ±2000 dps) ranges capture high-dynamic athletic motions without saturation.

SPI provides low-latency IMU-MCU transfer, and the IMU’s 512-byte FIFO supports burst reads and batching to reduce data loss at high sampling rates.

\begin{figure*}[!t]
    \centering
    \includegraphics[width=0.6\linewidth]{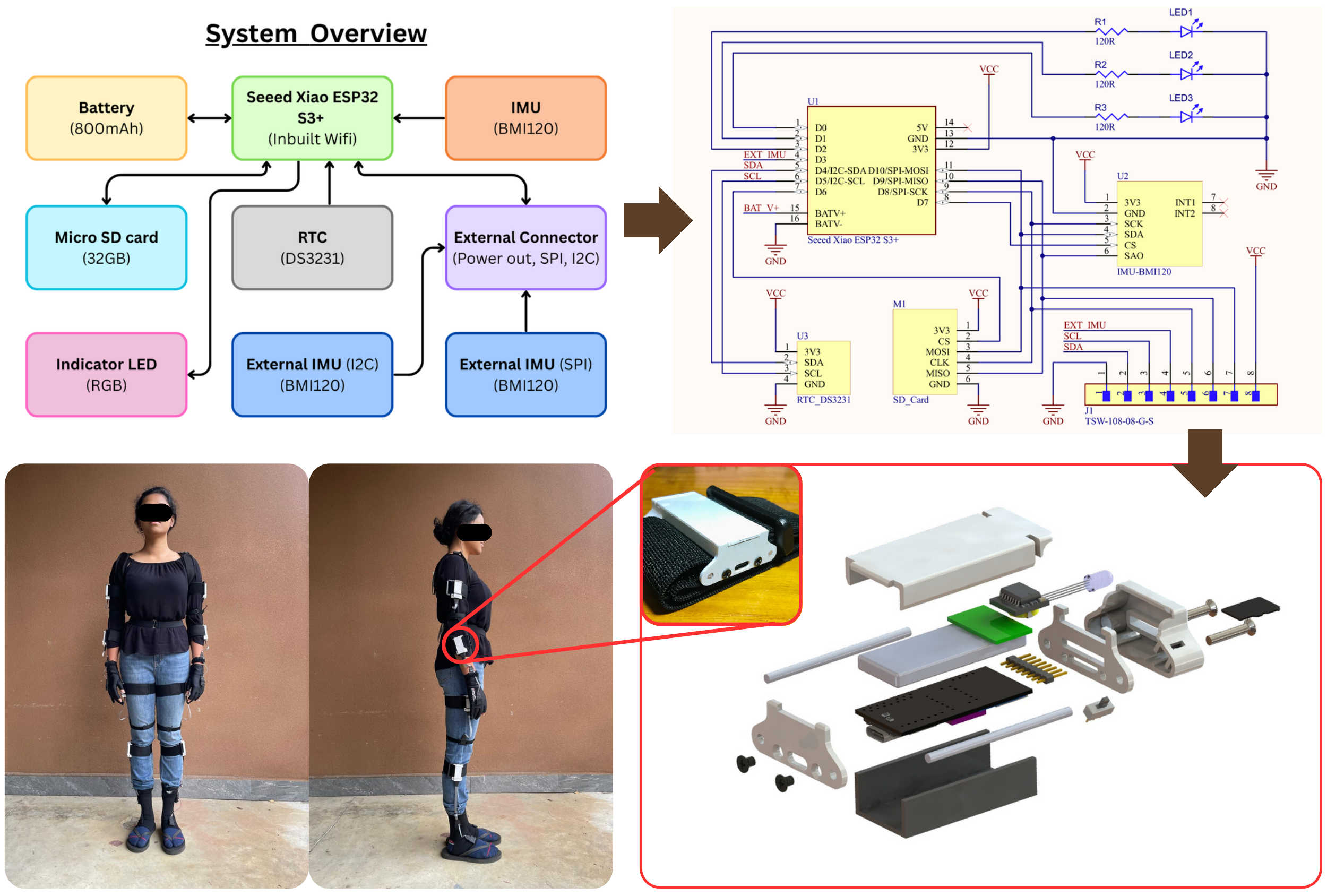}
    \caption{Custom 26-node IMU-based full-body motion capture system: system architecture overview showing key hardware components, (schematic circuit diagram, exploded view of the modular node enclosure, and system worn by a participant in a real-world configuration.}
    \label{fig:Diagram Demonstrating the entire device}
\end{figure*}

\subsubsection{Reliable Local Data Storage and Synchronization}

A microSD card module on the SPI bus provides non-volatile local logging to preserve data during temporary Wi-Fi interruptions. Each node uses a DS3231 RTC with a temperature-compensated crystal oscillator (TCXO) to generate stable timestamps and improve cross-node synchronization, supporting data integrity and temporal consistency across sessions and modules.

\subsubsection{Ergonomic Design and Power Management}

The device uses a 3.7 V, 800 mAh Li-Po battery for extended operation during active Wi-Fi streaming and recharging in approximately 6 hours. The enclosure uses a PLA body (RF transparent for wireless performance) with an aluminum backplate for rigidity and passive heat dissipation.

\subsubsection{Cost Benchmarking Against Commercial Solutions}

Whole device setup contains 10 main modules and each supports a maximum of two external IMUs to connect through 
its built-in connector, with the total sensor count scalable from 
1 to 30 to suit varying measurement requirements. The complete 
full-body setup costs approximately \$600~USD, significantly 
undercutting commercially available alternatives ranging from 
\$2{,}000~USD to \$30{,}000~USD~\cite{robertlachaine2020, adlou2025}, 
making this full-body IMU system a cost-effective solution.~\cite{choo2022, gonzalez2021, [R5]slade2021open}.

\subsection{Data Management and Software System Architecture}

\subsubsection{Real-Time operating system (RTOS)}

Firmware runs on FreeRTOS with priority based preemption. Sensor acquisition is prioritized to maintain consistent sampling, and SD logging is prioritized over TCP transmission to preserve data during transient link interruptions. FIFO queues decouple high-rate acquisition from slower consumer tasks, and mutexes prevent race conditions on shared resources. The integration of the TCP protocol is stable that enables wireless data transfer, guaranties stable delivery of packets, and sequenced communication between the device and the remote server. Overall, the architecture targets deterministic operation, high data integrity, and reliable real-time communication in multi-sensor deployments.

\begin{figure}[tb]
    \centering
    \includegraphics[width=1\linewidth]{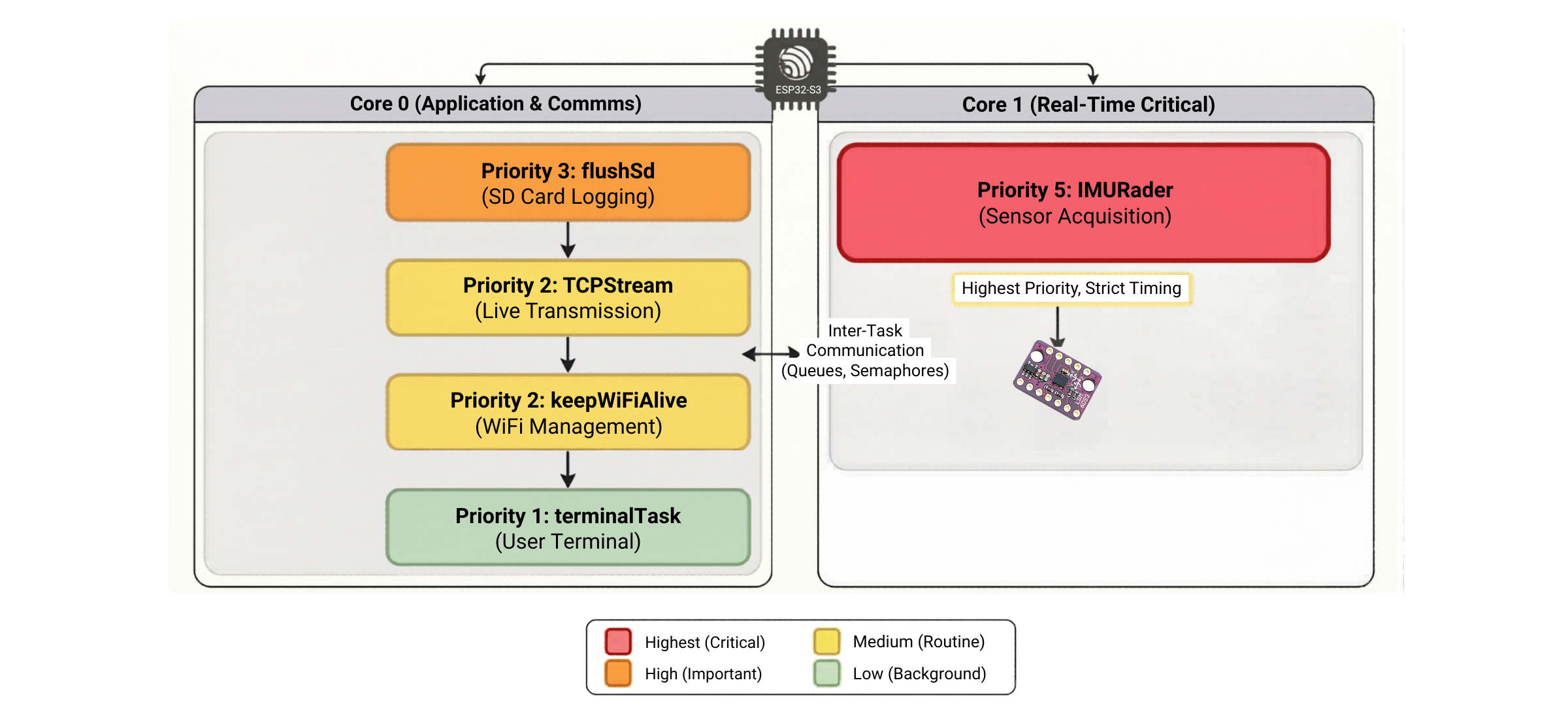}
    \caption{Dual-core firmware architecture of the ESP32-S3 microcontroller, illustrating FreeRTOS task prioritisation across Core 0 (Application and Communications) and Core 1 (Real-Time Critical sensor acquisition), with inter-task communication via queues and semaphores.}
    \label{fig:System Firmware Architecture}
\end{figure}

\subsubsection{High Fidelity Wireless Framework and Logging}

Local logging uses the SdFat library for fast SD access. To reduce non-deterministic flash delays, we use double-buffered, sector-aligned writes so one buffer is written while the other is filled, enabling continuous logging. For wireless streaming, TCP provides reliable, in-order delivery with error checking. A Batch and Timestamp Jitter Reduction Method assigns high-resolution timestamps at acquisition, buffers samples, and transmits uncoded batches; the receiver reconstructs the stream using only these trusted timestamps to achieve a jitter-reduced profile.

\subsubsection{Microsecond-Precision Time Synchronization with RTC and Network Time Stamps}

\begin{figure}[tb]
    \centering
    \includegraphics[width=0.9\linewidth]{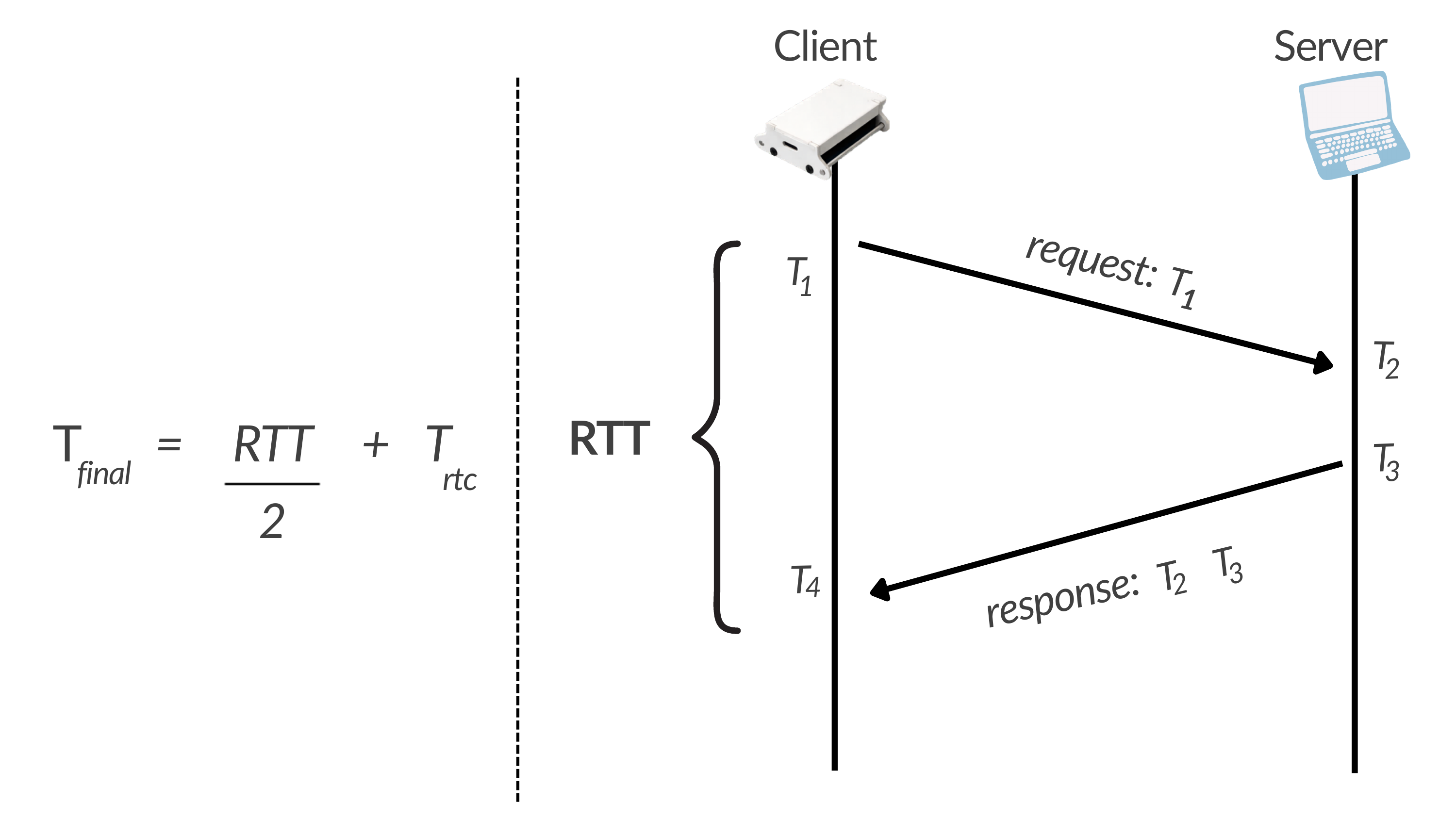}
    \caption{Four-timestamp synchronization protocol between the wearable IMU node (client) and server (host), used to estimate network round-trip latency and correct the onboard RTC and microsecond counter for sub-millisecond inter-node timing consistency.\cite{11499905}}
    \label{fig:sync_exchange}
\end{figure}

Fig.~\ref{fig:sync_exchange} illustrates the synchronization procedure. The server sends a reference time, while the node records local timestamps $t_1$ and $t_4$ and receives server timestamps $t_2$ and $t_3$. These four timestamps are used to estimate round-trip delay and correct the RTC plus the internal microsecond counter. This hybrid RTC-timestamp method reduces network-latency error and maintains sub-millisecond inter-node timing consistency.

\section{Measurement Model and Signal Processing Pipeline}
 \label{sec:pipeline}

This section describes the measurement model and the signal-processing pipeline implemented to estimate joint-angle trajectories from synchronized IMU recordings.
The pipeline converts raw 6 Degrees of Freedom (DoF) IMU streams (3-axis accelerometer and 3-axis gyroscope per sensor) into segment orientations using an indirect Kalman filter (IKF), computes relative segment rotation at each joint, extracts a scalar flexion/extension component, and applies drift mitigation and range normalization to produce the final output used throughout this manuscript.

\subsection{Coordinate Frames}
\label{sec:frames}

\begin{figure} [tpb]
    \centering
    \includegraphics[width=0.8\linewidth]{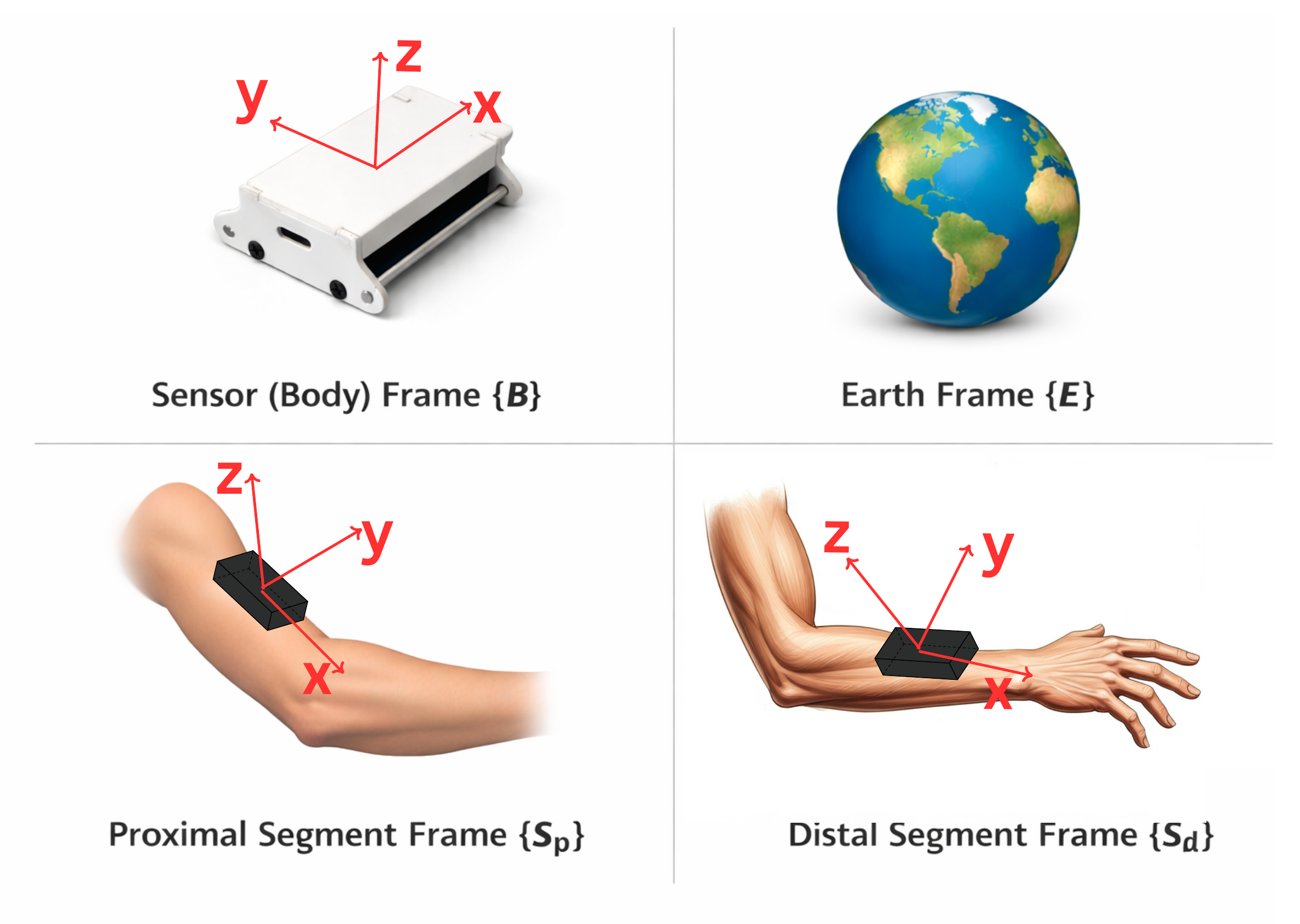}
    \caption{Coordinate frames considered. Sensor (body frame, Earth frame, and Segment frames (proximal and distal)}
    \label{fig: Coordinate frames.jpeg}
\end{figure}

As can be seen in Fig. \ref{fig: Coordinate frames.jpeg}, we define the following coordinate frames:
\begin{itemize}
    \item \textbf{Sensor (body) frame} $\{\mathcal{B}\}$: the local coordinate frame of each IMU (accelerometer/gyroscope axes).
    \item \textbf{Earth frame} $\{\mathcal{E}\}$: a gravity-referenced frame used for attitude representation and correction.
    \item \textbf{Segment frames} $\{\mathcal{S}_p\}$ and $\{\mathcal{S}_d\}$: proximal and distal segment orientations inferred from IMUs mounted on the corresponding segments.
\end{itemize}

\subsection{IMU Measurement Equations}
\label{sec:imu_model}

Each IMU provides a 3-axis angular rate measurement ($\boldsymbol{\omega}_{m,k}$) and a 3-axis specific force measurement ($\mathbf{a}_{m,k}$).
A standard discrete-time measurement model is:
\begin{align}
    \boldsymbol{\omega}_{m,k} &= \boldsymbol{\omega}_{k} + \mathbf{b}_{k} + \mathbf{n}_{\omega,k}, \label{eq:gyro_meas}\\
    \mathbf{a}_{m,k} &= \mathbf{C}_{k}^{\top}\big(\mathbf{a}_{g} + \mathbf{a}_{\mathrm{lin},k} + \mathbf{n}_{a,k}\big), \label{eq:acc_meas}
\end{align}
where $\boldsymbol{\omega}_{k}$ is the true angular rate in $\{\mathcal{B}\}$, $\mathbf{b}_{k}$ is gyroscope bias, $\mathbf{a}_{g}$ is gravitational acceleration in $\{\mathcal{E}\}$, $\mathbf{a}_{\mathrm{lin},k}$ is linear acceleration in $\{\mathcal{E}\}$, and $\mathbf{n}_{\omega,k}, \mathbf{n}_{a,k}$ denote measurement noise. Here, $\mathbf{C}_{k}$ is the rotation matrix.

% In preprocessing, accelerometer samples are normalized to units of $g$ (division by the constant $g$), and gyroscope samples are converted from degrees/s to radians/s prior to filtering.

\subsection{Assumptions}
\label{sec:assumptions}

The pipeline relies on the following assumptions:
\begin{itemize}
    \item \textbf{Rigid attachment:} each IMU is rigidly mounted to its corresponding segment.
    \item \textbf{Gravity-referenced correction:} accelerometer measurements provide a gravity direction reference (with possible contamination from linear acceleration), enabling drift reduction in the IKF.
    % \item \textbf{Dominant hinge component extraction:} joint flexion/extension is approximated by a selected Euler-angle component of the relative rotation $\mathbf{C}_{\mathrm{rel},k}$, using a fixed Euler sequence and a joint-dependent sign convention.
    \item \textbf{Uniform sampling:} processing assumes $T_s = 0.01$~s (100~Hz).
\end{itemize}

% \subsection{Assumptions}
% \label{sec:assumptions}

% The implemented pipeline relies on the following assumptions:
% \begin{itemize}
%     \item \textbf{Rigid attachment:} each IMU is rigidly attached to its corresponding body segment. 
%     % \item Any constant mounting offset is handled implicitly through initial offset removal and subsequent range mapping.
%     \item \textbf{Gravity-referenced correction:} accelerometer measurements provide a gravity direction reference (subject to contamination by linear acceleration during dynamic phases), enabling attitude drift reduction within the IKF.
%     \item \textbf{Dominant hinge component extraction:} joint flexion/extension is approximated by a selected Euler-angle component of the relative rotation $\mathbf{C}_{\mathrm{rel},k}$, using a fixed Euler sequence and a joint-dependent sign convention.
%     \item \textbf{Uniform sampling:} the processing assumes a sampling interval $T_s = 0.01$~s (100~Hz). 
%     % If a timestamp column is unavailable, a uniform time base is constructed as $t_k = k\,dt$.
% \end{itemize}

\subsection{Fusion Approach}
\label{sec:fusion}

Segment orientation for each IMU stream is estimated using an \emph{indirect Kalman filter} (IKF), which propagates attitude via gyroscope integration and corrects using the accelerometer-derived gravity direction while estimating gyroscope bias. At each sample, one IKF step outputs an updated quaternion, bias estimate, and rotation matrix $\mathbf{C}_{k}$. The IKF implementation follows Pandukabhaya et. al.\cite{[11]10858126}. The state and error-state are

\begin{equation}
\label{eq: state equation1}
\boldsymbol{x}_{k} =
\begin{bmatrix}
\hat{\underline{\boldsymbol{q}}}_{k} \\
\hat{\boldsymbol{b}}_{k}
\end{bmatrix}
\end{equation}

\begin{equation}
\label{eq: state equation2}
\tilde{\boldsymbol{x}}_{k} =
\begin{bmatrix}
\delta \boldsymbol{\theta}_{k} \\
\Delta \hat{\boldsymbol{b}}_{k}
\end{bmatrix}
\end{equation}

The continuous time state equations used are:
$$\mathbf{x}(t)=\begin{bmatrix}\bar{q}(t) \\ \mathbf{b}(t) \end{bmatrix}$$

\begin{align}
\dot{\bar{q}}(t) &= \frac{1}{2}\left(\bm{\omega}_m-\mathbf{b}-\mathbf{n_r}\right)\bar{q}(t) \\
\mathbf{\dot{b}} &= \mathbf{n_w}
\end{align}

Propagation and update equations are the same as those in M. Pandukabhaya et. al.\cite{[11]10858126}. Additionally, if an accelerometer sample is invalid (non-finite or near-zero norm), it is replaced by an initial gravity vector from the earliest valid sample to prevent filter instability.

\subsection{Joint Angle Estimation}
\label{sec:joint_angle_est}

Joint angles are estimated using one proximal and one distal IMU per joint. For each joint and time index $k$:

\paragraph*{Relative rotation}
At time $k$, segment orientation is represented by a rotation matrix $\mathbf{C}_{k}\in\mathbb{R}^{3\times 3}$. $\mathbf{C}_{k}$ is a proper rotation (orthonormal with $\det(\mathbf{C}_{k})=+1$). The proximal-to-distal relative rotation is
\begin{equation}
    \label{eq:rel_rot}
    \mathbf{C}_{\mathrm{rel},k} = \mathbf{C}_{p,k}^{\top}\mathbf{C}_{d,k}.
\end{equation}
From $\mathbf{C}_{\mathrm{rel},k}$, a scalar flexion component is extracted. In the validation and case study(as shown in sections \ref{sec:validation_standardized}, \ref{sec: sensitivity, resolution, drift}, \ref{sec:protocol_cj}, and \ref{sec:perf_cj}) this component is one \texttt{xyz} Euler component, converted to degrees, and multiplied by a joint-specific sign to enforce consistent flexion direction. Or else, a dimension reduction method such as PCA can be used to obtain a specific dominant rotation component. The resulting sequence is wrapped to $(-180^\circ,180^\circ]$.

\paragraph*{Phase unwrapping}
The wrapped sequence is unwrapped to obtain a continuous flexion signal:
\begin{equation}
    \label{eq:unwrap}
    \tilde{\theta}_{k} = \mathrm{unwrap}\big(\theta^{\mathrm{wrap}}_{k}\big)
\end{equation}

\paragraph*{Initial offset removal}
An initial window of $N_0$ samples is used to remove a constant offset. Absorbs fixed mounting offsets and the initial pose reference; optional and used when needed:
\begin{equation}
    \label{eq:init_offset}
    \theta^{\mathrm{bend}}_{k} = \tilde{\theta}_{k} - \frac{1}{N_0}\sum_{j=1}^{N_0}\tilde{\theta}_{j}
\end{equation}

\paragraph*{Mapping to an anatomical-like angle convention}
A configurable base angle maps the bend signal to an anatomical-like angle:
\begin{equation}
    \label{eq:base_map}
    \theta^{\mathrm{raw}}_{k} = \theta_{\mathrm{base}} - \theta^{\mathrm{bend}}_{k}
\end{equation}

\paragraph*{Drift mitigation}
The integration of gyroscope bias through the IKF results in a slow drift being accumulated overtime. To attenuate slow drift, while preserving rapid movement components, the pipeline applies a Butterworth high-pass filter to the unwrapped flexion sequence (order 2, cutoff $f_c=0.05$~Hz), producing $\tilde{\theta}^{\mathrm{HP}}_{k}$. The value of $f_c=0.05$~Hz was selected by sweeping $f_c$ from 0.01 Hz to 1 Hz to find an optimal mitigation of drift. The high-pass filter was applied in an offline post-processing manner using zero-phase forward–backward filtering. The same offset removal and mapping in \eqref{eq:init_offset}-\eqref{eq:base_map} are then applied to obtain an intermediate high-pass angle.
A constant shift $s$ then maps the waveform into a target physiological range (default $[40^\circ,175^\circ]$) without changing its shape:
\begin{equation}
    \label{eq:range_shift}
    \theta^{\mathrm{final}}_{k} = \tilde{\theta}^{\mathrm{HP}}_{k} - s.
\end{equation}
In the implementation, $s$ is chosen from the extrema-based lower/upper bound shifts (using their midpoint), with a feasibility flag indicating whether a shift-only operation can satisfy both bounds; if not, the midpoint shift is still applied (optionally followed by clipping). The resulting $\theta^{\mathrm{final}}_{k}$ is the primary joint-angle estimate used in this manuscript.

\begin{figure} [tpb]
    \centering
    \includegraphics[width=0.88\linewidth]{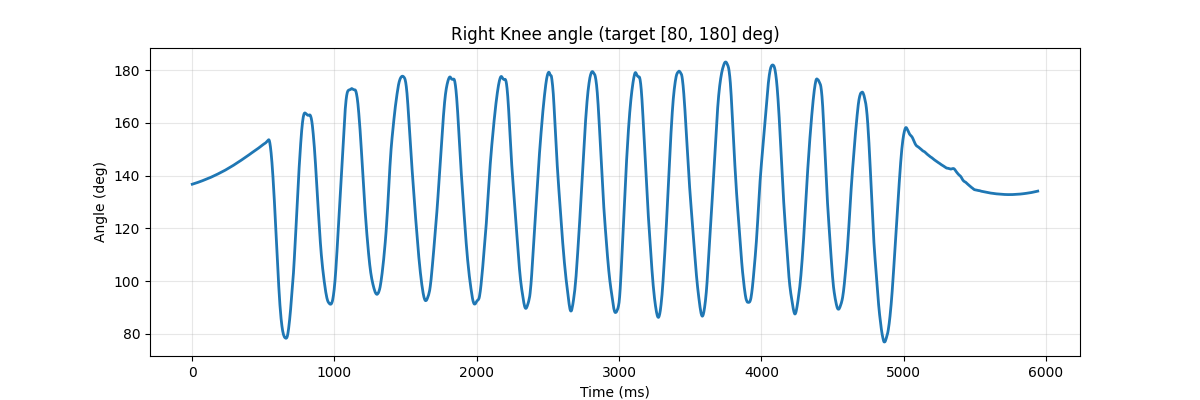}
    \caption{Right knee angle variation extracted from IMU data using high pass + range shift method}
    \label{fig: Right_knee_HPF_After.png}
\end{figure}

\begin{figure} [tpb]
    \centering
    \includegraphics[width=0.8\linewidth]{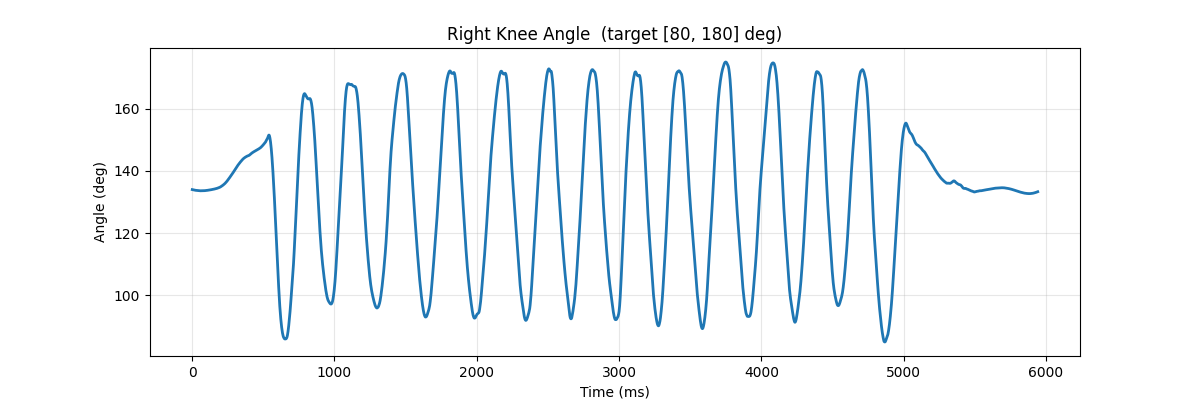}
    \caption{Right knee angle variation extracted from IMU data using KL/SSA method}
    \label{fig: Right_knee_KL_After.png}
\end{figure}

\paragraph*{Validation using KL Transform}
For diagnostic comparison, a Karhunen-Lo\`eve/Singular Spectrum Analysis (KL/SSA) detrending path is also implemented. The agreement between the KL/SSA result (Fig.~\ref{fig: Right_knee_KL_After.png}) and the high-pass-based estimate (Fig.~\ref{fig: Right_knee_HPF_After.png}) serves as an internal consistency check that the dominant motion structure is preserved under two drift-mitigation approaches. Reference-based \emph{validation} against an external standard is reported separately in Section~\ref{sec:validation_standardized}.

\section{Reference-Based Device Validation Using Standardized Motion}
\label{sec:validation_standardized}

This section presents the reference-based validation of the proposed device and signal-processing pipeline. The validation was performed against a computer-vision-based reference in order to assess whether the wearable system can reliably capture the temporal variation and waveform structure of joint-angle motion. The objective of this validation is therefore not to establish absolute-angle accuracy, but to evaluate shape-based agreement, including the preservation of cycle structure, phase behaviour, and dominant kinematic patterns in the measured joint-angle trajectory.

\subsection{Experimental Setup and Standardized Motion Protocol}
\label{sec:validation_setup}

\begin{figure} [tpb]
    \centering
    \includegraphics[width=0.6\linewidth]{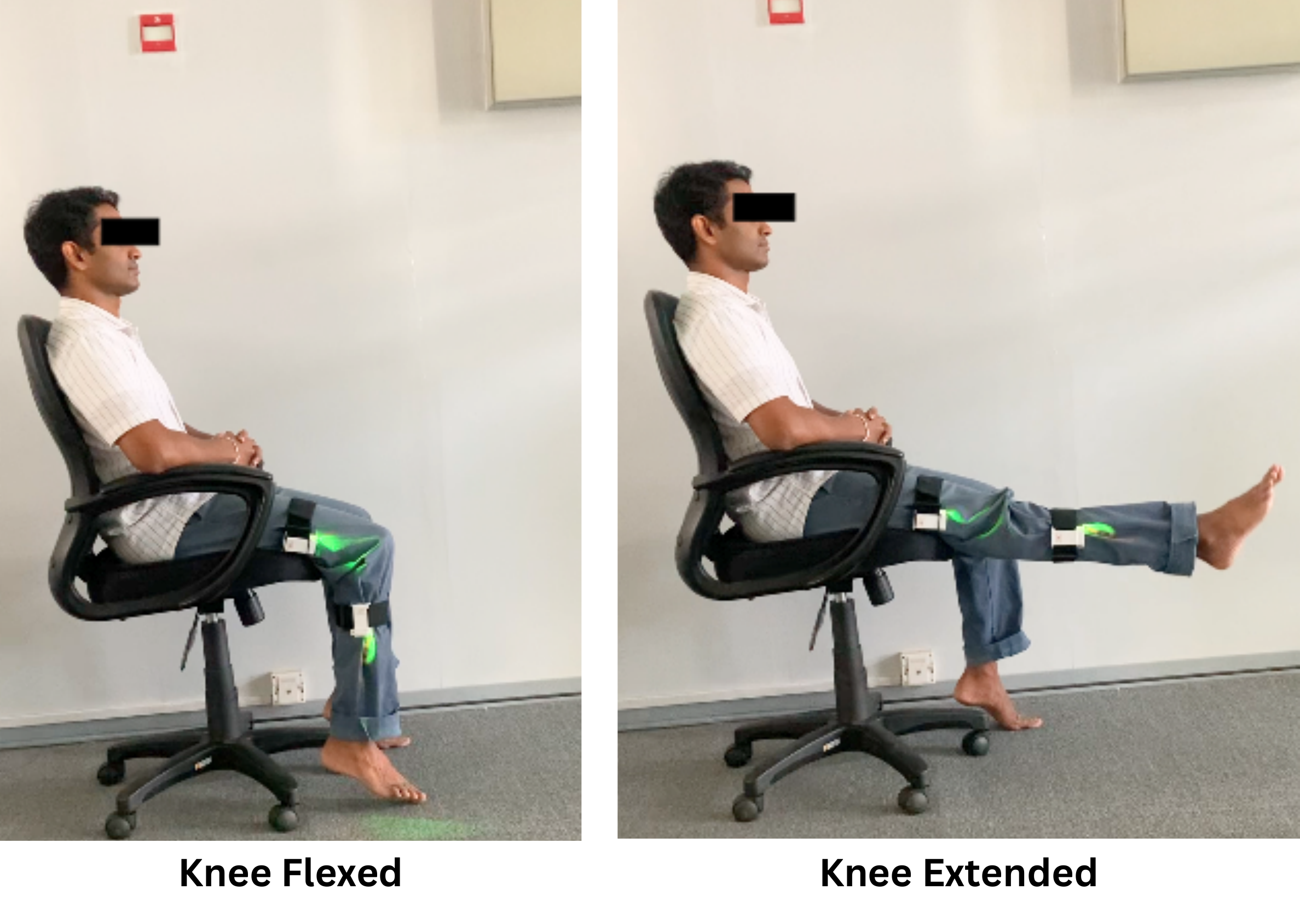}
    \caption{Device Validation through Knee Flexion and Extension. Images are frames of video captured at an angle perpendicular to the plane of motion}
    \label{Knee flexion and extension.png}
\end{figure}

A standardized right-knee flexion-extension protocol (Fig.~\ref{Knee flexion and extension.png}) was used to validate the IMU-based wearable pipeline against a vision-derived reference.
The subject, seated with both feet on the floor, repeatedly extended and flexed the right leg for 14 cycles.
The motion was predominantly sagittal-plane, with the camera approximately perpendicular to the plane of motion.

Video was recorded at $30$~fps, while inertial data were collected from thigh (proximal) and shank (distal) IMUs at $\Delta t = 0.01$~s (100~Hz).
The objective is to compare the wearable knee-angle time series with the knee angle computed from video-based pose estimation during this controlled single-joint motion.

% \label{sec:validation_standardized}

% \subsection{Experimental Setup and Standardized Motion Protocol}
% \label{sec:validation_setup}

% \begin{figure} [tpb]
%     \includegraphics[width=1\linewidth]{Images/Knee flexion and extension.png}
%     \caption{Device Validation through Knee Flexion and Extension. Images are frames of video captured at an angle perpendicular to the plane of motion}
%     \label{Knee flexion and extension.png}
% \end{figure}

% A standardized right-knee flexion-extension protocol as shown in fig. \ref{Knee flexion and extension.png} was conducted to validate the proposed IMU-based wearable instrumentation and processing pipeline against a vision-derived reference.
% The subject sat with both feet on the floor and repeatedly extended the right leg to near full extension before returning to the flexed position, completing 14 flexion–extension cycles.
% The motion was performed predominantly in a sagittal plane, with the camera positioned approximately perpendicular to the plane of motion.

% A monocular video was recorded at $30$~fps.
% In parallel, inertial data were acquired from two wearable IMU nodes mounted on the thigh (proximal segment) and shank (distal segment) at a sampling interval of $\Delta t = 0.01$~s (100~Hz).
% The objective of this experiment is to compare the knee-angle time series estimated by the wearable system with the knee-angle time series computed from video-based pose estimation during a controlled cyclic single-joint movement.

\subsection{Reference Signal Extraction Using Video-Based Pose Estimation}
\label{sec:validation_reference}

The video reference was computed using a YOLOv11 pose-estimation model \cite{[14]Redmon2016YOLO}, which outputs 2D pixel coordinates and confidence values for each detected keypoint per frame \cite{[13]UltralyticsYOLO11Docs}.
For each frame $k$, the right hip, right knee, and right ankle keypoints were extracted, and frames with low keypoint confidence were removed using a threshold $\tau_c = 0.5$ applied to all three joints.

The right-knee angle was then computed in the image plane as the inter-segment angle between the thigh and shank vectors:
\begin{equation}
\label{eq:cv_angle}
\begin{aligned}
\mathbf{v}_{1}(k) &= \mathbf{p}_{\mathrm{hip}}(k)-\mathbf{p}_{\mathrm{knee}}(k) \\
\mathbf{v}_{2}(k) &= -\mathbf{p}_{\mathrm{ankle}}(k)-\mathbf{p}_{\mathrm{knee}}(k) \\
\theta_{\mathrm{CV}}(k)&=\cos^{-1}\!\left( \frac{ (\mathbf{v}_{1}(k))^\top
(\mathbf{v}_{2}(k))
}{
\|\mathbf{v}_{1}(k)\|
\;
\|\mathbf{v}_{2}(k)\|}\right)
\end{aligned}
\end{equation}
where $\mathbf{p}_{\mathrm{hip}}(k)$, $\mathbf{p}_{\mathrm{knee}}(k)$, and $\mathbf{p}_{\mathrm{ankle}}(k)$ denote 2D pixel coordinates of the corresponding keypoints in frame $k$.
Angles were converted to degrees.
Missing frames due to confidence gating were filled by linear interpolation, and the resulting trajectory was smoothed using a Savitzky-Golay filter (window length $11$ frames, polynomial order $3$).

\paragraph*{Reference uncertainty considerations}
The vision-derived knee angle $\theta_{\mathrm{CV}}$ is a 2D image-plane quantity and is sensitive to out-of-plane motion, camera perspective/parallax, and keypoint noise, particularly near full extension. Accordingly, $\theta_{\mathrm{CV}}$ is treated as a practical reference for approximately planar motion rather than an error-free ground truth.

\subsection{Quantitative Comparison Metrics}
\label{sec:validation_metrics}

\paragraph*{Normalization}
Since the IMU and video derived knee angles span different ranges, both trajectories were max-normalized to $[0,1]$ to enable shape-based comparison.

\paragraph*{Time-base alignment}
The IMU and video signals are sampled at different rates (100~Hz vs 30~Hz). One trajectory is resampled to the other's time base using linear interpolation. A residual temporal offset $\Delta t_0$ between $\theta_{\mathrm{IMU}}$ and $\theta_{\mathrm{CV}}$ is estimated using cross-correlation (or peak alignment) and compensated prior to error computation.

\paragraph*{Error metrics}
After alignment, the instantaneous error is
\begin{equation}
\label{eq:error}
e(k) = \theta_{\mathrm{IMU,norm}}(k) - \theta_{\mathrm{CV,norm}}(k),
\end{equation}
and we report RMSE, MAE and Bias as shown in section \ref{sec:metrics};
% \begin{align}
% \label{eq:rmse}
% \mathrm{RMSE} &= \sqrt{\frac{1}{K}\sum_{k=1}^{K} e(k)^2},\\
% \label{eq:mae}
% \mathrm{MAE}  &= \frac{1}{K}\sum_{k=1}^{K} |e(k)|,\\
% \label{eq:bias}
% \mathrm{Bias} &= \frac{1}{K}\sum_{k=1}^{K} e(k),
% \end{align}
where $K$ is the number of aligned samples. 

\paragraph*{Check for Qualitative Agreement}
Aligned trajectories are also inspected for agreement in onset/offset, number of cycles, and phase. Systematic differences are interpreted in the context of reference uncertainty and measurand definition.

\begin{figure} [tpb]
    \centering
    \includegraphics[width=0.8\linewidth]{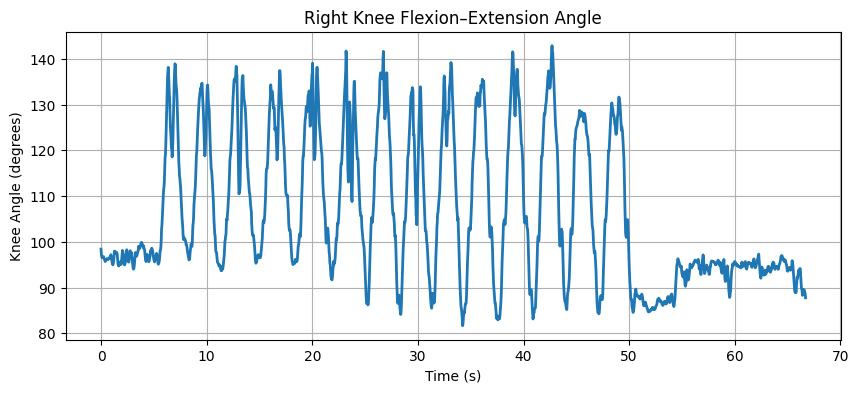}
    \caption{Vision-derived right-knee angle $\theta_{\mathrm{CV}}(t)$ computed from YOLOv11 2D keypoints (hip-knee-ankle) during the standardized seated flexion--extension trial.}
    \label{fig: Right_knee_CV Result.png}
\end{figure}

\subsection{Validation Results and Interpretation}
\label{sec:val_results}

All IMU results correspond to the final high-pass + range-shift estimate shown in Fig. \ref{fig: Right_knee_HPF_After.png}. These results were compared to the computer vision output as below.

\paragraph*{Trajectory comparison (shape agreement).}
Fig. ~\ref{fig: Right_knee_CV Result.png} and Fig. ~\ref{fig: Right_knee_HPF_After.png} show the vision-derived and IMU-derived knee-angle trajectories for the 14-cycle trial. When observed side-by-side, the agreement in the dominant waveform shape and cycle structure can be seen; indicating that the wearable pipeline captures the primary kinematic signature of the standardized motion.

\paragraph*{Quantitative error summary}
Table~\ref{tab:val_metrics} summarizes RMSE, MAE, and bias after normalization and time based alignment. 

\begin{table}[t]
    \centering
    \caption{Summary of vision-reference validation metrics for standardized knee flexion--extension.}
    \label{tab:val_metrics}
    \begin{tabular}{lcc}
        \toprule
        \textbf{Metric} & \textbf{Value} & \textbf{Notes} \\
        \midrule
        RMSE & 0.2515 & over aligned interval \\
        MAE  & 0.2231 & over aligned interval \\
        Bias & -0.0386 & mean($\theta_{\mathrm{IMU}}-\theta_{\mathrm{CV}}$) \\
        \bottomrule
    \end{tabular}
\end{table}

\paragraph*{Sources of discrepancy (interpretation of error)}
Differences are expected due to (i) limitations of the vision reference and (ii) modality-specific measurand definitions. The sources for discrepancies are as follows:
\begin{itemize}
    \item \textbf{2D projection effects:} Out-of-plane motion and perspective/parallax can bias $\theta_{\mathrm{CV}}$, especially near full extension.
    \item \textbf{Keypoint noise:} Residual localization error may perturb extrema despite confidence gating and smoothing (Section~\ref{sec:validation_reference}).
    \item \textbf{Different measurand definitions (2D vs 3D definition):} The reference uses a 2D inter-segment angle, while the wearable uses a 3D relative rotation with one Euler component, so axis/alignment differences can shift amplitudes.
    % \item \textbf{Dynamic acceleration:} Linear acceleration can affect gravity-based correction in the IKF during faster segments.
    \item \textbf{Sampling rate/alignment limits:} Resampling and residual offsets (video-frame/IMU-sample scale) can introduce local phase error in steep-slope regions.
\end{itemize}

\paragraph*{Validation conclusion}
Overall, the standardized-motion experiment shows that the wearable pipeline captures the expected knee flexion-extension cycle structure with low normalized error. Remaining discrepancies are consistent with markerless 2D reference limitations and measurand differences, supporting use of the IMU output for clean \& jerk characterization (Section~\ref{sec:perf_cj}).

\section{Instrumentation Performance of the Device and Algorithm}
\label{sec: sensitivity, resolution, drift}

This section analyzes the sensing performance of the device through its resolution, while also investigating drift mitigation effectiveness. The evaluation is carried out using a long-duration rigid-body hold experiment; where the elbow of the \textbf{rigid body} was constrained at approximately \textbf{$115^\circ$ for 2~hours and 12~minutes}.

\subsection{Resolution}
\label{sec:sens_resolution}

\paragraph*{Noise-limited practical resolution}
To quantify a practical (noise-limited) resolution, we used the above mentioned long duration hold test. Once the elbow angle was obtained, the standard deviation ($\sigma$) of the elbow angle for the entire 2 hours and 12 minutes was calculated. Using that, the practical resolution ($\delta\theta_{\mathrm{res}}$) was calculated through equation \ref{eq:practical_resolution}.

We define the \emph{practical resolution} as a $2\sigma$ noise band,
\begin{equation}
\label{eq:practical_resolution}
\delta\theta_{\mathrm{res}} \triangleq 2\,\sigma_{\mathrm{pooled}}
\end{equation}
This follows the common metrological use of a $k=2$ coverage factor, where $\pm 2\sigma$ gives approximately 95\% coverage under near-normal scatter \cite{NISTTN1297Expanded, NISTCoverageFactor, NISTHandbookExpandedUnc}. 
% The difference between $\sigma_1$ and $\sigma_2$ suggests non-identical noise across windows (e.g., minor disturbances or settling), while the pooled estimate provides a conservative single-number summary.

Thus, as $\sigma = 0.3221^\circ$, $\delta\theta_{\mathrm{res}} = 0.6442^\circ$.
In sport-specific validations of wearable IMU kinematics, typical joint-angle errors are in the order of several degrees \cite{Nijmeijer2023Xsens}. For example, in tennis, tolerable angle error magnitudes are around $5-10 ^\circ$ for elbow angles \cite{RuizMalagon2022NOTCH}, reinforcing that our $\delta\theta_{\mathrm{res}}$ is far below typical sports measurement errors and therefore supports detecting subtle technique-related changes.

% \paragraph*{Sensitivity}
% Here, \emph{sensitivity} is the \textbf{\textit{change in reported joint angle per change in true joint angle}} (static gain). Since the pipeline extracts a dominant hinge-motion coordinate from relative segment rotation and applies a final range mapping, the nominal sensitivity is unity. Future work will quantify sensitivity using known step/ramp angle changes (jig/goniometer) via linear fits (slope = sensitivity, intercept = bias), or regression against an external reference trajectory.

%------------------------OLD----------------------------------
% \subsection{Sensitivity and Resolution}
% \label{sec:cj_sensitivity}

% Sensitivity and resolution were characterized using the noise floor observed in quasi-static periods (e.g., base posture segments) and the smallest reliably detectable within-lift changes.
% For each joint, the angular noise standard deviation $\sigma_{\theta}$ was computed over a static window, and the practical resolution was defined as
% \begin{equation}
% \theta_{\mathrm{res}} \approx 2\sigma_{\theta},
% \end{equation}
% corresponding to a conservative two-sigma detectability threshold.
% The measured dynamic trajectories show that the system resolves both large joint excursions and finer technique-dependent modulations (e.g., the jerk dip depth and elbow stabilization fluctuations), indicating sufficient sensitivity for posture-feature extraction during high-dynamic tasks.

% -----------Insturctions-----------------
% TODO (authors): Insert $\sigma_{\theta}$ and $\theta_{\mathrm{res}}$ values (deg) for 1--2 representative joints.

\subsection{Long-duration Stability and Drift Accumulation}
\label{sec:drift}

As explained in section \ref{sec:joint_angle_est}, we use a high-pass filter based method to address the issue of slow drift accumulation. This drift mitigation provides long-duration stability for the device to measure joint angles continuously for a long period of time, which is very important in sports-like dynamic motion analysis. 
The long-duration stability was evaluated using quasi-static baseline windows at the beginning and end of the long-duration hold test described above. The initial and final baselines were defined as the first and last 150~s of the recording, respectively.

Let $\theta(t)$ denote the measured elbow-angle trajectory. We computed the baseline means $\bar{\theta}_{\mathrm{initial}}$ and $\bar{\theta}_{\mathrm{final}}$ (and corresponding standard deviations) over the initial and final windows. Drift was summarized by the baseline mean shift
\begin{equation}
\Delta \theta_{\mathrm{drift}} = \bar{\theta}_{\mathrm{initial}} - \bar{\theta}_{\mathrm{final}},
\end{equation}
and the time-normalized drift rate
\begin{equation}
r_{\mathrm{drift}} = \frac{\Delta \theta_{\mathrm{drift}}}{T/60}\quad \text{(deg/min)},
\end{equation}
where $T$ is the total recording duration in seconds.

The results yielded $\bar{\theta}_{\mathrm{initial}} = 115.702^\circ$ and $\bar{\theta}_{\mathrm{final}} = 115.702^\circ$, corresponding to $\Delta\theta_{\mathrm{drift}} = (1.99 \mathrm{e}{-4}) ^\circ$ and $r_{\mathrm{drift}} = 1.5 \mathrm{e}{-6}\,\text{deg/min}$. These values indicate near-complete suppression of long-term drift over multi-hour durations, supporting the suitability of the proposed processing chain for long recordings without requiring re-zeroing.

\section{Experimental Protocol for Dynamic Task - Clean and Jerk}
\label{sec:protocol_cj}

This section describes the experimental protocol used to evaluate the proposed wearable joint-angle instrumentation under a highly dynamic, multi-joint movement: the clean and jerk (C\&J).
The C\&J was selected because it includes rapid transitions across multiple posture states, large joint excursions, and high angular velocities, thereby stressing both the sensing hardware (sampling, timing consistency) and the signal-processing pipeline.

\subsection{Participants}
Two participants were recruited to represent distinct skill levels:
(i) a professional, national-level weightlifter and (ii) an amateur weightlifter. The demographic and expertise level details of each lifter is shown in table \ref{tab:participant_demographics}.

\begin{table}[t]
    \centering
    \caption{Participant demographics and expertise level}
    \label{tab:participant_demographics}
    \begin{tabular}{lllll}
        \toprule
        \textbf{Participant}& \textbf{Expertise} & \textbf{Sex} & \textbf{Age (yrs)} & \textbf{Height / Mass} \\
        \midrule
        Professional & National-level & \textit{Female}& \textit{23}& \textit{160 cm} / \textit{49 kg}\\
        Amateur & Amateur athlete & \textit{Male}& \textit{21}& \textit{166 cm} / \textit{61 kg} \\
        \bottomrule
    \end{tabular}
\end{table}

\noindent\textbf{Ethics and consent:}
This study was approved by the Ethical Review Committee of the Faculty of Engineering, University of Peradeniya, Sri Lanka (Approval No.ERC/FoE/2025/001), and written informed consent was obtained from all participants.

% TODO (authors): Insert participant demographics (age, height, weight, sex) for each participant.
% Example: "Participant P1 (professional): age=..., height=..., weight=..., sex=...; Participant P2 (amateur): ..."
%
% TODO (authors): Insert ethics statement (IRB/ethics approval identifier, institution, and informed consent procedure).

\subsection{Task Definition and Trial Structure}
Each participant performed the clean and jerk with a $20$~kg barbell, while the wearable recorded at $100$~Hz. Each participant completed two trials, with five consecutive lifts per trial, performed at a self-selected comfortable pace.
%
% TODO (authors): If available, specify rest duration between trials and/or between lifts,
% and report the approximate duration of each trial (e.g., "each trial lasted approximately ... s").

\subsection{Trial Selection for Analysis}
Only one trial per participant was retained for most of the analysis. The retained trial was selected as the \emph{best} of the two based on clear completion of all five lifts, reducing bias from incomplete motion sequences in subsequent characterization and stage-wise comparisons.

\subsection{Clean and Jerk Stages}
\begin{figure}[t]
    \centering
    \includegraphics[width = 0.7\linewidth]{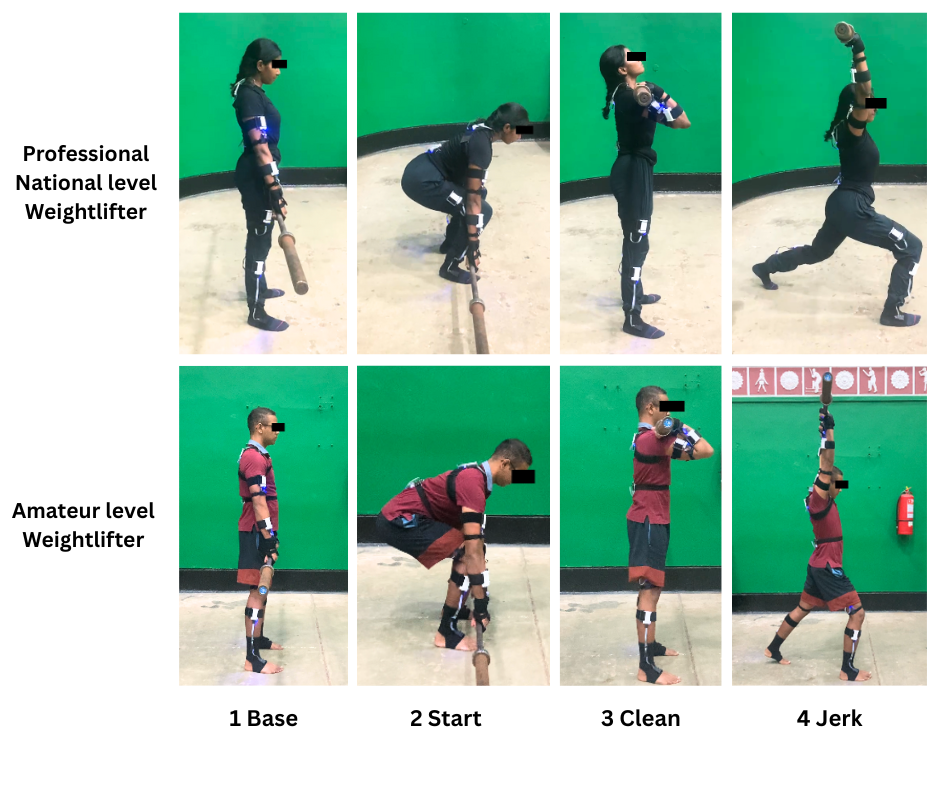}
    \caption{Stages of the clean and jerk lift used for the dynamic-task protocol: (1) base, (2) start, (3) clean, and (4) jerk.}
    \label{fig:cj_stages}
\end{figure}

For interpretability, each lift was segmented into four canonical stages (Fig.~\ref{fig:cj_stages}):
\begin{enumerate}
    \item \textbf{Base:} initial upright posture with the barbell held prior to initiating the lift.
    \item \textbf{Start:} descent and setup phase leading into the explosive pull.
    \item \textbf{Clean:} acceleration and receipt of the barbell at the shoulders (rack position).
    \item \textbf{Jerk:} dip-and-drive and overhead stabilization.
\end{enumerate}
These stages enable consistent lift alignment and stage-wise joint-angle comparison.

%  Old version

% Data were collected from 10 participants performing standardized clean and jerk motions using an empty Olympic barbell (20 kg). The wearable motion capture device comprising 26 wireless IMUs was worn throughout the experiment, with sensor placement as illustrated in Fix X. All sensors were sampled at 100 Hz and synchronized using the proposed communication framework.

% Each trial consisted of 5–6 consecutive clean and jerk lifts, following a standardized sequence: the barbell was initially held stationary, a complete clean and jerk lift was performed, the bar was held briefly in a static position (< 1 s), returned to the starting position, and held stationary again for approximately 2 s before the next repetition. This protocol was designed to introduce repeatable dynamic and quasi-static phases for subsequent measurement analysis. Video data were simultaneously recorded from a camera positioned approximately perpendicular to the participant’s right sagittal plane to support as reference during analysis.

\section{Instrument Utility through a Dynamic Task - Clean and Jerk}
\label{sec:perf_cj}

This section characterizes utility of the wearable device and joint angle measuring pipeline through a dynamic exercise, which is the clean and jerk (C\&J) in weightlifting. Unless otherwise stated, all joint angles are the final high-pass + range-shift output, and results are discussed in terms of signal shape (exact values can be recovered by scaling and shifting).

\subsection{Stage (of lift) Segmentation Enabled by Measurements}
\label{sec:cj_stage_seg}

\begin{figure}[tpb]
   \centering
    \includegraphics[width=0.5\textwidth]{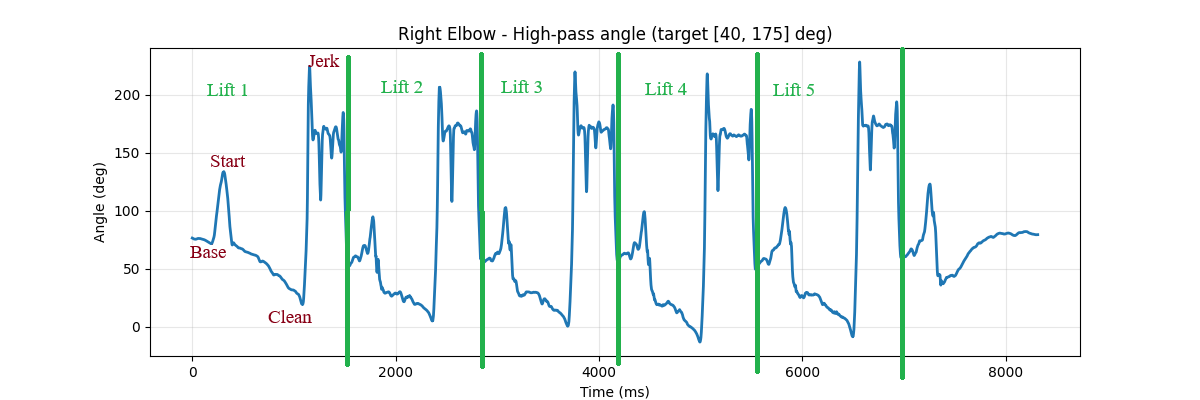}
   \caption{Variation of the right elbow angle of professional lifter segmented to 5 lifts and 4 stages of a lift (Base, Start, Clean and Jerk)}
   \label{fig: Professional Elbow right new segmented}
\end{figure}    

The C\&J stage structure is visible in the measured elbow and knee trajectories. That is to say that our device and algorithmic pipeline enables the differentiation of the four stages of each lift and five lifts of each trial. Each stage in Fig.~\ref{fig:cj_stages} produces a repeatable signature: \textbf{base} is quasi-static, \textbf{start} shows a knee-flexion trough, \textbf{clean} contains steep coordinated transitions, and \textbf{jerk} exhibits a knee dip-and-drive with an elbow extension/plateau. Preserving these spike-like transients is an instrumentation outcome of the wearable measurements and processing chain, enabling stage-wise interpretation. An example using the right elbow is shown in Fig.~\ref{fig: Professional Elbow right new segmented}.

\subsection{Inter-trial Variability}
\label{sec:cj_repeatability}

To assess inter-trial variability of the professional weightlifter, 2 methods of comparisons were done for the 2 trials of (C\&J) lifts. The 2 methods were the comparison of probability (to compare the probability distribution of joint angles of the 2 trials) and comparison of waveform similarity (to compare the temporal waveform via cross correlation)
% Is the above description good enough?

\subsubsection*{Comparison of probability}
\label{sec:cj_kldivergence}

Inter-trial consistency was evaluated by comparing empirical distributions of joint-angle trajectories for the right/left elbows and knees. For each joint, normalized histograms were computed using common binning for both trials, yielding distributions $P$ (Trial~1) and $Q$ (Trial~2). Discrepancy was quantified using KL divergence \cite{KL_div}.
\begin{equation} 
D_{\mathrm{KL}}(P\|Q)=\sum_i P_i \log \frac{P_i}{Q_i},
\end{equation}
and the symmetric form
\begin{equation}
D_{\mathrm{SKL}}(P,Q)=\frac{1}{2}\left[D_{\mathrm{KL}}(P\|Q)+D_{\mathrm{KL}}(Q\|P)\right].
\end{equation}
Lower values indicate greater similarity; histogram correlation was also computed as a complementary measure.

As summarized in Table~\ref{tab:inter_trial_distributional_variability}, the right knee showed the strongest inter-trial consistency (lowest symmetric KL and highest correlation), while the left knee exhibited the greatest divergence. Elbow joints showed intermediate consistency, with a noticeable shift in mean angles between trials (e.g., right elbow: $79.259^\circ$ to $59.511^\circ$). These results reflect both measurement variability and genuine trial-to-trial movement variation.

\begin{table}[t]
    \centering
    \caption{Inter-trial distributional variability of joint-angle trajectories for the professional participant (two C\&J trials)}
    \label{tab:inter_trial_distributional_variability}
    \begin{tabular}{lcccc}
        \toprule
        \textbf{Joint} & Right elbow & Left elbow & Right knee & Left knee \\ 
        \midrule
        $\boldsymbol{D_{\mathrm{KL}}(T1\|T2)}$ & 0.939 & 1.223 & 0.398 & 1.146 \\
        $\boldsymbol{D_{\mathrm{KL}}(T2\|T1)}$ & 1.580 & 1.972 & 1.854 & 2.819 \\
        \textbf{Symmetric KL} & 1.259 & 1.598 & 1.126 & 1.983 \\
        % \textbf{Hist. Corr.} & 0.189 & 0.223 & 0.450 & 0.438 \\
        % $\boldsymbol{\mu_{T1}}$ ($^\circ$) & 79.259 & 96.807 & 123.459 & 127.495 \\
        % $\boldsymbol{\mu_{T2}}$ ($^\circ$) & 59.511 & 78.761 & 130.195 & 107.157 \\        
        \bottomrule
    \end{tabular}
\end{table}

\subsubsection*{Comparison of waveform similarity (Cross-Correlation)}
\label{sec:cj_crosscorr}

Cross-correlation was used to assess waveform similarity between Trial~1 and Trial~2. For each joint, the normalized cross-correlation was computed over a range of lags, and the maximum correlation coefficient ($r_{\max}$) and its lag were extracted. Higher $r_{\max}$ indicates greater shape similarity, while smaller peak lag indicates stronger temporal alignment.

Table~\ref{tab:crosscorr_summary} summarizes the cross-correlation results. The right and left elbows showed the highest inter-trial similarity, while the left knee exhibited the lowest $r_{\max}$ and a comparatively larger lag, consistent with greater variability in stabilization dynamics. Overall, the cross-correlation results agree with the distributional analysis in indicating higher repeatability for the upper-limb trajectories than for the left-knee trajectory across trials.

\begin{table}[t]
\centering
\caption{Summary of inter-trial cross-correlation between Trial~1 and Trial~2 joint-angle trajectories.}
\label{tab:crosscorr_summary}
\begin{tabular}{lcc}
\toprule
\textbf{Joint} & $\boldsymbol{r_{\max}}$ & \textbf{Peak Lag (samples)} \\
\midrule
Right elbow & 0.6782& 187\\
Left elbow  & 0.6604& 152\\
Right knee  & 0.5436& 151\\
Left knee   & 0.3283& 183\\
\bottomrule
\end{tabular}
\end{table}

\subsection{Demonstration of Expertise-level Discrimination by Observing Measurement-derived}
\label{sec:cj_expertise}

\begin{figure}[tpb]
   \centering
    \includegraphics[width=0.4\textwidth]{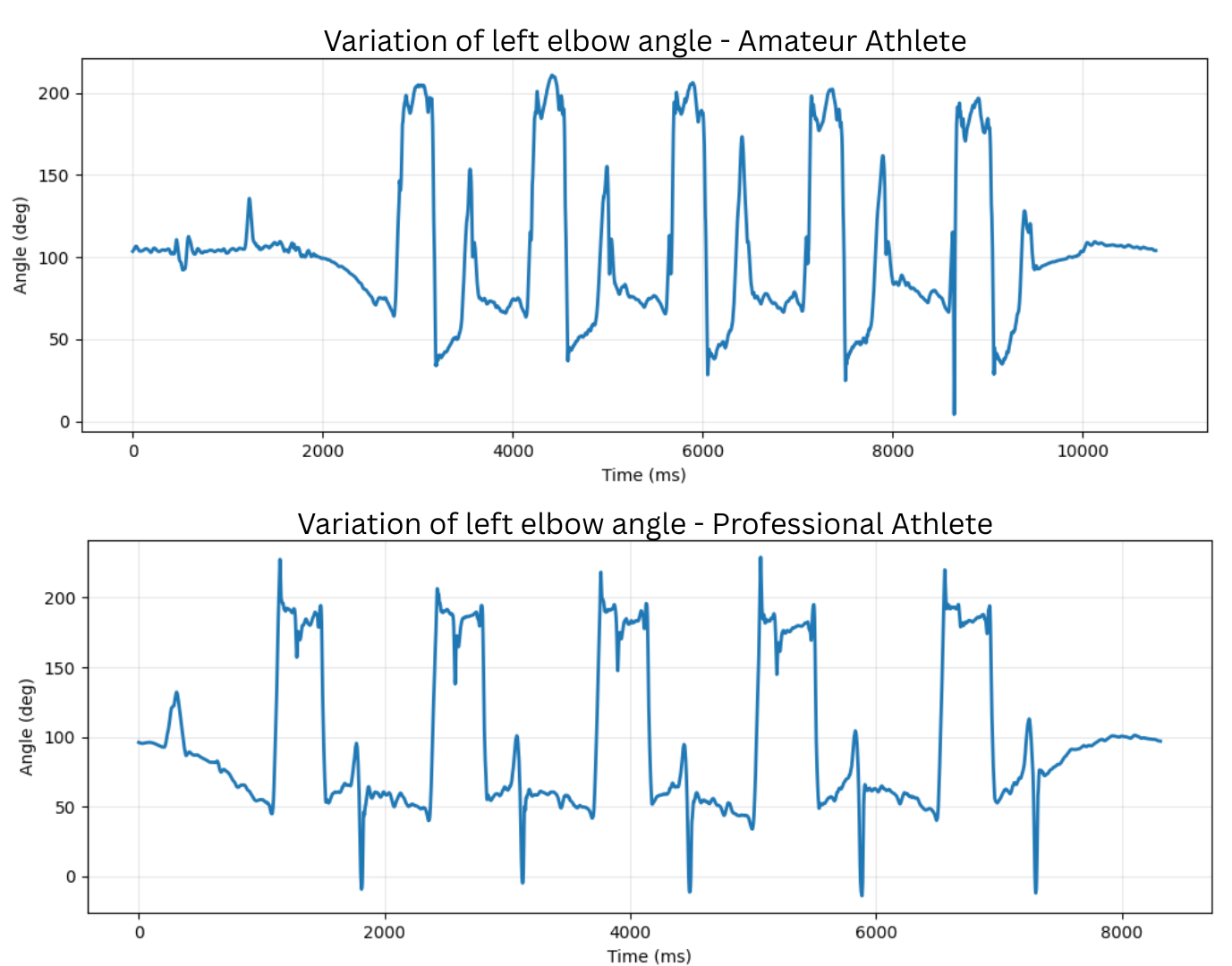}
   \caption{Expertise level discrimination. The jerky nature of the amateur athlete can be clearly seen as compared to the professional athlete. The controlled nature of the professional athlete allows for more explosive movements when necessary. These can be clearly observed from the results.}
   \label{fig: Expertise level discrimination}
\end{figure}

Beyond stability and repeatability, the wearable outputs are sensitive enough to reveal technique-dependent differences between the professional and amateur participants during the C\&J (Fig.~\ref{fig: Expertise level discrimination}).

\paragraph*{Explosive movement signatures}
Explosiveness appears as brief, high-slope transients in the joint-angle trajectories. For the professional participant, these transients occur mainly in the expected lift phases, while the remaining phases stay controlled. In contrast, the amateur participant show weaker (lower slope) transients when rapid acceleration is expected and occasional abrupt changes outside those phases, indicating less effective timing of forceful actions.

\paragraph*{Smooth versus jerky motion}
Smooth technique appears as regular, repeatable waveforms with little high-frequency fluctuation, whereas jerky execution shows visible jitter on the main trajectory. The professional participant exhibits smoother and more repeatable elbow and knee trajectories, while the amateur participant shows greater irregularity and within-lift variability. Overall, preserving these signatures while mitigating slow drift highlights a key instrumentation outcome: the system enables technique-sensitive discrimination directly from the measured joint-angle signals.

\section{Conclusion}
\label{sec:conclusion}

This paper presented a synchronized multi-IMU wearable instrumentation platform and processing pipeline for multi-joint angle measurement in standardized and high-dynamic sports movements. The system integrates a low-cost, high-density 26-node design with robust local logging, an RTC-disciplined microsecond timestamping strategy for sustained inter-node temporal consistency, an IKF-based orientation estimator,  relative-rotation for joint-angle extraction, with drift mitigation through high-pass filtering, and range normalization.

Reference-based validation used a standardized seated knee flexion--extension protocol with a markerless YOLOv11 vision reference, demonstrating close agreement in cycle structure over 14 repetitions after time-base alignment and max normalization for shape comparison. Instrumentation performance was then characterized through a long-duration rigid-body hold test (2~h 12~min), which demonstrated near-complete drift suppression ($r_{\mathrm{drift}}=1.5\times10^{-6}$ deg/min) and sub-degree practical resolution ($\delta\theta_{\mathrm{res}}=0.6442^\circ$), supporting long-duration monitoring without re-zeroing. Finally, the system’s utility was demonstrated on a C\&J task using a professional and an amateur participant performing five consecutive lifts per trial, where the measured elbow and knee trajectories preserved stage-dependent signatures and fast transients enabling stage-wise interpretation, inter-trial variability analysis, and expertise-level discrimination.

% Future work will (i) validate against higher-accuracy references (e.g.: marker-based motion capture or instrumented hinge/goniometric rigs) to quantify absolute-angle accuracy without normalization, (ii) formalize synchronization evaluation using larger event sets and controlled benchmarks, and (iii) expand evaluation to larger cohorts and broader sports movements while accounting for alignment uncertainty, soft-tissue artifact, and high-acceleration effects on gravity-based correction. 

The functionality of this device will be further generalized as a precise joint angle measurement system by calibration against higher accuracy references and incorporating other sports.

Overall, the proposed hardware + algorithm wearable system provides a practical, scalable, and generalizable approach for temporally consistent multi-joint posture estimation and technique analysis in real-world sports settings.

\bibliographystyle{IEEEtran}
\bibliography{references}

\end{document}